\documentclass[12pt]{article}
\usepackage[totalheight=23cm, totalwidth = 17cm]{geometry}
\newcommand{\nin}{\noindent}
\newcommand{\be}{\begin{equation}}
\newcommand{\ee}{\end{equation}}
\newcommand{\bea}{\begin{eqnarray}}
\newcommand{\eea}{\end{eqnarray}}

\newcommand{\nn}{\nonumber\\}

\begin{document}
\begin{titlepage}

\begin{center}
{\Large {\bf Antisymmetric-Tensor and Electromagnetic effects
in an $\alpha^{'}$-non-perturbative Four-Dimensional String Cosmology}}

\vspace{1.5cm}

{\bf Jean Alexandre, Nick E. Mavromatos} and {\bf Dylan Tanner} \\
\vspace{0.2cm}
{\it King's College London, Department of Physics, Strand, London WC2R 2LS, U.K.}
\vspace{2cm}

{\bf Abstract}

\end{center}

\nin
Starting from an exact (in the Regge slope $\alpha^{'}$) functional method  for a bosonic stringy $\sigma$-model,
we investigate four-dimensional cosmological string solutions in graviton, dilaton and antisymmetric tensor backgrounds,
compatible with world-sheet conformal invariance, and valid beyond perturbative expansions in powers of $\alpha^{'}$.
The antisymmetric tensor field, playing
the r\^ole of an axion in the four-dimensional target space time, leads to spatial anisotropies of the
emergent Robertson-Walker expanding Universe, and, upon coupling the system to the electromagnetic
field, it results in non-trivial optical activity. Some estimates of the corresponding effects are made and
their relevance to current cosmology is briefly discussed.

\vspace{8cm}
\begin{flushleft}
August 2007
\end{flushleft}

\end{titlepage}

\section{Introduction}

The usual approach to  (low-energy) string cosmology~\cite{ABEN1,ABEN2,copeland2,stringcosmo} proceeds via
the construction and study of effective actions, which are perturbative in the Regge slope $\alpha^{'}$, thereby employing a target-space derivative expansion scheme. The actions are constructed by the requirement that the respective equations of motion correspond to the vanishing of
the appropriate  renormalization-group beta functions,
or more precisely
the Weyl-anomaly coefficients~\cite{metsaev},
for the various (cosmological) string backgrounds of the pertinent stringy $\sigma$-model. The Weyl-anomaly coefficients are obtained by the usual renormalization procedure of canceling
short-distance world-sheet divergences, leading to renormalized $\sigma$-model couplings/background fields. In a cosmological setting, these background fields depend only on the time coordinate $X^0$, as appropriate for the description of a homogeneous cosmology to a first approximation. This amounts to solving a set of differential equations, which are usually truncated to the first few orders (most commonly up to quartic) in an expansion in powers of $\alpha^{'}$, under the assumption that for late eras of the Universe the space-time curvature is small, such that only the first few orders of curvature tensors in the gravitational part of the effective action play a significant r\^ole in the underlying physics.

However, such perturbative derivative-expansion treatments may not be appropriate for a discussion of the physics of phase transitions in the Early Universe, where one may face highly-curved situations.
To this end, an alternative approach was developed in \cite{AEM1}, where a new configuration
of time-dependent (cosmological) stringy graviton and dilaton backgrounds was derived, satisfying Weyl invariance in a
$\alpha^{'}$-non-perturbative way.

The key point to achieve such configurations was to obtain beta functions which are homogeneous in the time $X^0$, to all orders in the Regge slope
$\alpha^{'}$, thereby depending only on overall constants. The next step was to perform a reparametrization of the string (that is, appropriate field redefinitions of the various background fields)~\cite{metsaev} in order to fine tune these overall constants to zero. From an effective field theory viewpoint,  these field-redefinitions leave the scattering amplitudes of the target space invariant, and therefore there are known as ``field redefinition ambiguities'', in the sense that not all of the coefficients in the expressions of the Weyl anomaly coefficients, and thus the associated terms in the target-space effective action, can be fixed by the (perturbative) physics~\footnote{It should be noted that such ambiguities may be lifted by some truly non-perturbative target-space symmetries, for instance $O(d,d)$ space-time invariance~\cite{meissner}, characteristic of some cosmological string models~\cite{oddsymm}. Our approach depends crucially on the freedom to be able to perform string background reparametrizations, which in turn implies that the emergent cosmologies might not be characterized by such symmetries. As we do not know the $\alpha^{'}$-non-perturbative target-space-time action in a closed form, the issue of such invariances is at present unclear to us, and will not be attempted to be discussed here.}.

For the
configuration found in \cite{AEM1}, this reparametrization actually does not change the $X^0$ dependence of the
backgrounds. It was then argued that the above-mentioned fine tuning can always be performed, whatever the
target space dimension is. In this way, exact (all orders in $\alpha^{'}$)
target-space cosmological configurations have been found, consistent with conformal invariance, and thus acceptable  space-time backgrounds of strings.
Upon restriction to four target-space dimensions, one recovers an Einstein-frame Minkowski Universe, with a logarithmic (in cosmic-time)
dilaton field.

It should be remarked at this point that,
within a Wilsonian world-sheet renormalization-group framework, which notably is different from the method developed in \cite{AEM1},
the above configurations correspond to  infrared fixed points. In this sense,
such Universes might be considered as exit phases from an expanding Universe in string theory, e.g. the linearly expanding Universe of \cite{ABEN1,ABEN2}, which notably corresponds to a {\it trivial} ``fixed'' point of the ``flow'' with respect to the parameter that controls world-sheet quantum fluctuations in the approach of \cite{AEM1}. The Minkowski space-time with a logarithmic dilaton, on the other hand, corresponds to a {\it non-trivial} ``fixed'' point of that ``flow''.

We propose here to include the Kalb-Ramond antisymmetric tensor field of the string spectrum and perform, in a similar way as in \cite{AEM1}, the corresponding study of the bosonic $\sigma$-model. We shall also analyze briefly the physical consequences of the emergent
time-dependent metric of the associated homogeneous Cosmology.
The reader should recall that, in four space time dimensions, the Kalb-Ramond field strength, $H_{\rho\mu\nu}$, is dual to an axion-like (pseudoscalar) field~\cite{ABEN1,ABEN2}, which is related
\cite{battye} to the Peccei-Quinn phase transition \cite{peccei}. The axion can also be related to the
quintessence field \cite{choi}, and also to cosmic optical activity, associated with a large-scale anisotropy
in the propagation of electromagnetic fields in the Universe~\cite{birefringence}.
The antisymmetric tensor usually implies an anisotropic metric \cite{copeland1}, but it is also possible to
consider an isotropic metric, either with an axion depending on time only~\cite{ABEN1,ABEN2},
or with a plane wave axion, after a space-averaging of the energy
momentum tensor corresponding to $H_{\rho\mu\nu}$ \cite{copeland2}.

In this work
we will consider a generalization of the approach of \cite{AEM1},
to include such antisymmetric-tensor-induced axion-like target-space background fields. In our approach,
instead of demanding the order-by-order-in-$\alpha^{'}$ vanishing of the world-sheet beta functions in order to achieve perturbative
Weyl invariance,
we will impose the requirement that the Weyl anomaly coefficients
of the various backgrounds
are homogeneous functions of $X^0$, to all orders in $\alpha^{'}$.
In this context, then, we shall demonstrate that, in the framework of a four-dimensional target space, it is possible to find
a non-trivial $X^0$-dependence for the antisymmetric tensor field strength,
$H_{\mu\nu\rho}$,
provided a spatial direction is singled out in its definition.
This results in an non-trivial optical activity, when the field strength $H_{\rho\mu\nu}$ is
augmented by an Abelian gauge field~\cite{birefringence},
as appropriate for string theory.

The article is structured as follows: In section 2, we present an exact functional method to study the quantum version of a bosonic $\sigma$-model in (homogeneous cosmological) graviton, dilaton and Kalb-Ramond antisymmetric tensor backgrounds, by means of which we derive the evolution
of the dilaton field with $\alpha^{'}$. The latter parameter controls the magnitude of quantum fluctuations, and thus plays the r\^ole of the ``control parameter'' of the approach~\cite{AEM1}. In this way, we obtain the evolution of the dilaton with the amplitude
of the (world-sheet) quantum fluctuations. We demonstrate that it is possible to obtain a condition on the configuration of the background
metric and dilaton fields for which the dilaton is independent of the amplitude of quantum fluctuations. The antisymmetric tensor affects
the dilaton evolution indirectly, through the evolution of the metric tensor.
We then use this condition, in section 3, to study the world-sheet conformal properties of such a configuration,
and find that a specific power law (with respect to the time $X^0$) for the space-time metric leads to
world-sheet beta functions which are homogeneous in $X^0$. We explain how to achieve world-sheet conformal invariance by means of
an appropriate set of field redefinitions for the various background fields (``reparametrization of the string''), which does not change the functional dependence of the
backgrounds, but affects the beta functions in  such a way that it is always possible to make them vanish (in some renormalization scheme).
Notably, this study in done is
four dimensional target space time, which is compatible with world-sheet conformal invariance of the string, thus alleviating the need for compactification. In section 4, we discuss the cosmological relevance of this configuration. We find that the direction singled out by the
antisymmetric tensor field is linearly expanding with the Einstein-frame time, whereas the remaining two spatial directions can
be static, for a specific choice of the dilaton amplitude, thereby leading to an anisotropically expanding Universe. In section 5 we study
the coupling of the antisymmetric tensor field
to an Abelian (electromagnetic) gauge field, in a way specific to string models. We restrict our analysis to lowest order in an $\alpha^{'}$ expansion,
where the (bosonic) string effective action exhibits a universal structure among string theories. We prove the Weyl invariance of our solution, upon the inclusion of the back reaction effects of the electromagnetic field onto space time, to this order and present arguments for the validity of this result to all orders in $\alpha^{'}$. We
discuss briefly the phenomenological implications of such a coupling, namely the resulting
optical activity of the Universe, implying the rotation of the direction of the electric field with the cosmic time.
This, in principle, can lead to  measurable effects. Finally, section 6 contains our conclusions and prospects for future studies. Some technical aspects of our approach are presented in two Appendices.

\section{Control of quantum fluctuations and $\lambda$-flows}

We commence our discussion by considering a spherical world sheet with a Euclidean metric and curvature scalar $R^{(2)}$.
The bare action of a $\sigma$ model describing the propagation of a first-quantized bosonic string in graviton and dilaton and Kalb-Ramond antisymmetric tensor
backgrounds reads:
\be
S=\frac{1}{4\pi}\int d^2\xi\sqrt{\gamma}\left\{\lambda\left(\gamma^{ab}\eta_{\mu\nu}
+\varepsilon^{ab}a_{\mu\nu}\right)\partial_a X^\mu\partial_b X^\nu
+R^{(2)}\phi_{bare}(X^0)\right\},
\ee
where $\eta_{\mu\nu}$ is the flat Minkowski target-space metric, and $a_{\mu\nu}$
is a coordinate-independent bare antisymmetric tensor, which, as we shall demonstrate in this work, will introduce an anisotropy in the expansion of the Universe. Motivated by homogeneous cosmology,
we assume that the  bare dilaton field $\phi_{bare}$ is a function of the time coordinate only.
The parameter $\lambda$ plays, therefore, the r\^ole of $1/\alpha^{'}$ and can be used~\cite{AEM1} to
control the magnitude of the quantum fluctuations,
which are proportional to $\alpha^{'}$. It parametrizes the quantum theory
described by the effective action, i.e., the proper-graph-generating functional $\Gamma_\lambda$,
which is defined in Appendix A. The range of values for $\lambda$ is $[1/\alpha^{'};\infty[$:
\begin{itemize}
\item $\lambda\to\infty$ corresponds to $\alpha^{'}\to 0$ and therefore to a classical theory:
the kinetic term dominates over the bare dilaton $\phi_{bare}$ term, and the theory is free;
\item $\lambda\to 1/\alpha^{'}$ generates the full quantum theory: the interactions arising from $\phi_{bare}$ are
gradually switched on as $\lambda$ decreases from $\infty$ to $1/\alpha^{'}$.
\end{itemize}
In this work we seek a $\lambda$-independent configuration of the bosonic string, which, by construction, is non-perturbative since it is independent of the strength of $\alpha^{'}$.

\vspace{.5cm}

We derive in Appendix A the exact equation describing the evolution
of  the proper-graph-generating functional
$\Gamma_\lambda$
with respect to the parameter $\lambda$, which reads
\bea\label{evolG}
\dot\Gamma_\lambda&=&
\frac{1}{4\pi}\int d^2\xi\sqrt{\gamma}\left(\gamma^{ab}\eta_{\mu\nu}+\varepsilon^{ab}a_{\mu\nu}\right)
\partial_a X^\mu\partial_b X^\nu\nn
&&+\frac{1}{4\pi}\mbox{Tr}\left\{\left(\gamma^{ab}\eta_{\mu\nu}+\varepsilon^{ab}a_{\mu\nu}\right)
\frac{\partial}{\partial\xi^a}\frac{\partial}{\partial\zeta^b}
\left(\frac{\delta^2\Gamma_\lambda}{\delta X^\mu(\zeta)\delta X^\nu(\xi)}\right)^{-1}\right\}
\eea
where the overdot denotes derivative with respect to $\lambda$.
In eq.(\ref{evolG}), the symbol of the trace (defined in eq.(\ref{trace}) of Appendix A)
contains the quantum corrections to $S$. In order to obtain physical information on the system,
eq.(\ref{evolG}) should be integrated from $\lambda=\infty$ to $\lambda=1/\alpha^{'}$,
which is the appropriate regime of the full quantum theory.

An important remark is in order, to avoid confusion: {\it unlike} a Wilsonian approach, where a running regulator
would be used for the trace appearing in eq.(\ref{evolG}),
and $\lambda=1/\alpha^{'}$ would be fixed, we use here a {\it fixed} regulator and let $\lambda$ run.
The bare kinetic term therefore {\it does not} play the r\^ole of a regulator, and the trace in eq.(\ref{evolG})
will be regularized by a {\it fixed} ultraviolet world sheet cut-off $\Lambda$, which is necessary in order
to define properly our flow with respect to $\lambda$.

Anticipating the r\^ole of the antisymmetric tensor in inducing anisotropies in the Universe's expansion, we consider a general, anisotropic, configuration for the space-time metric:
\be
g_{\mu\nu}=\mbox{diag}\left(\kappa(X^0),\tau_1(X^0), ..., \tau_{D-1}(X^0)\right).
\ee
In order to solve the evolution equation (\ref{evolG}), we need to assume a functional dependence for the
effective action. In the framework of the gradient expansion, we consider the following ansatz:
\bea\label{gradexp}
\Gamma_\lambda&=&\frac{1}{4\pi}
\int d^2\xi\sqrt{\gamma}\left\{\gamma^{ab}\kappa(X^0)\partial_a X^0\partial_bX^0
+\gamma^{ab}\sum_{i=1}^{D-1}\tau_i(X^0)\partial_a X^i\partial_b X^i\right.  \nn
&& ~~~~~~~~~~~~~~~+\varepsilon^{ab}B_{\mu\nu}(X^0)\partial_a X^\mu\partial_b X^\nu+R^{(2)}\phi(X^0)\Bigg\},
\eea
where $\kappa,\tau_i,\phi$ are $\lambda$-dependent functions of $X^0$

We consider the limit where the radius of the world sheet goes to infinity, while keeping the curvature
scalar $R^{(2)}$ finite.
We show in Appendix A that the Ansatz (\ref{gradexp}), when plugged into the exact
evolution equation (\ref{evolG}), leads to:
\be\label{equa1}
\dot\phi=-\frac{\Lambda^2}{2R^{(2)}}\left(\frac{1}{\kappa}+\sum_{i=1}^{D-1}\frac{1}{\tau_i}\right)+
\frac{\phi^{''}}{4\kappa^2}\ln\left(1+\frac{2\Lambda^2\kappa}{R^{(2)}\phi^{''}}\right)\nonumber,
\ee
where $\Lambda$ is the world-sheet ultraviolet cut-off,
and the prime denotes derivative with respect to $X^0$. Note that the antisymmetric tensor does not
appear in the evolution for the dilaton. As explained in Appendix A, the evolution equation for the dilaton
is obtained by taking a constant configuration for the fields $X^\mu$, for which off diagonal
terms (in the space-time indices) do not play any r\^ole. The antisymmetric tensor will play a r\^ole in the
evolution equation for the metric, which can be obtained by considering non-constant configurations. Therefore
the evolution of the dilaton is indirectly coupled to the antisymmetric tensor, via the metric. The evolution
of the latter does not need to be derived explicitly, though, as it will be obtained by means of the world-sheet Weyl invariance conditions.

\vspace{0.5cm}

A $\lambda$-flow fixed point solution $\dot\phi=0$ is possible for $\kappa=F\phi^{''}$ (with $F$ a constant):
\be
\kappa\dot\phi=-\frac{\Lambda^2}{2R^{(2)}}\left(1+\kappa\sum_{i=1}^{D-1}\frac{1}{\tau_i}\right)
+\frac{1}{4F}\ln\left(1+\frac{2\Lambda^2F}{R^{(2)}}\right),
\ee
provided that the following additional condition is satisfied
\be
\kappa\sum_{i=1}^{D-1}\frac{1}{\tau_i}=-1+\frac{R^{(2)}}{2\Lambda^2F}
\ln\left(1+\frac{2\Lambda^2F}{R^{(2)}}\right)
=\mbox{constant}.
\ee
The above constant is negative, as required for a Minkowski space-time signature, for a sufficiently large cut-off $\Lambda$.
In order to obtain an $\alpha^{'}$-independent solution, one must satisfy then the following conditions:
\bea\label{conditions}
\kappa(X^0)&\propto&\phi^{''}(X^0)\nn
\sum_{i=1}^{D-1}\frac{1}{\tau_i(X^0)}&\propto&\frac{1}{\kappa(X^0)}.
\eea
These conditions are independent of world-sheet Weyl-invariance properties of the background fields. The latter should be imposed as extra constraints. We shall do so in the next section, where we will discuss
Weyl invariance for a
specific {\it four-dimensional} background-field configuration
satisfying the conditions (\ref{conditions}), which thus will play the r\^ole of a consistent string ground state.

\section{Weyl Invariance  properties (in D=4 target space-time dimensions) of the $\lambda$-flow fixed point}

It was shown in \cite{AEM1} that the $\alpha^{'}$-non-perturbative configuration of graviton and dilaton backgrounds, which appears as a fixed point of the corresponding $\lambda$-flow evolution equation (\ref{evolG}),
satisfies local world-sheet conformal invariance in
any target space-time dimension. We consider here the extension of such an analysis for the case where a Kalb-Ramond antisymmetric
tensor field in dimension $D=4$ is included.
Such an extension can be motivated phenomenologically, as a consequence
of the coupling of the antisymmetric tensor field with electromagnetic fields in string theory, resulting in non-trivial optical activity in the
resulting Universe~\cite{birefringence}, as we shall discuss in section 5.

The most general
$X^0$-dependent expression for the antisymmetric tensor field strength $H_{\rho\mu\nu}$ is
\be
H_{\rho\mu\nu}=\omega~\varepsilon_{\rho\mu\nu\sigma}\partial^\sigma h,
\ee
where:
\begin{itemize}
\item $\omega$ is a function of $X^0$ only;
\item $\varepsilon_{\rho\mu\nu\sigma}$ is the totally antisymmetric Levi-Civita tensor density, with $\varepsilon^{\mu\nu\rho\sigma} = |g|^{-1} \varepsilon_{\mu\nu\rho\sigma}$ and
$\varepsilon_{0123}=\sqrt{|g|}=\sqrt{|\kappa\tau_1\tau_2\tau_3|}$;
\item $h$ is a function linear in space directions and independent of time.
\end{itemize}
Note that, when one solves one-(world-sheet)-loop (${\cal O}(\alpha^{'})$) equations of motion for $H_{\rho\mu\nu}$, $\omega=\exp(4\phi)$
and $h$ is identified as the four-dimensional axion (pseudoscalar) field~\cite{ABEN1}.
In our case, though, as we do not impose the vanishing of the one-loop beta functions,
$\omega$ is independent of the dilaton field. We will show that the
choice
\bea
h&=&h_0X^3\nn
\omega&=&1,
\eea
where $h_0$ a constant, is consistent with Weyl invariance in a $\alpha^{'}$-non-perturbative manner.
This simple choice suffices for our purposes, since,
as we shall see, it encompasses
the
essential features of our approach: it leads to $\alpha^{'}$-non-perturbative background configurations, and also
implies a non-trivial optical activity in the
Universe, as we shall discuss in section 5.
A more general expression for $H_{\rho\mu\nu}$ will be studied in a forthcoming publication, which will
include a complete study of the different possibilities that might arise.

\subsection{First order in $\alpha^{'}$}

To first order in $\alpha^{'}$, the beta functions for the bosonic world-sheet
$\sigma$-model theory in graviton, antisymmetric tensor and dilaton backgrounds are \cite{metsaev}:
\bea\label{weylconditions}
\beta_{\mu\nu}^{g(1)}&=&R_{\mu\nu}+2\nabla_\mu\nabla_\nu\phi-\frac{1}{4}H_{\mu\rho\sigma}H_\nu^{~\rho\sigma}\nn
\beta_{\mu\nu}^{B(1)}&=&-\frac{1}{2}\nabla^\rho H_{\rho\mu\nu}+\partial^\rho\phi H_{\rho\mu\nu}\nn
\beta^{\phi(1)}&=&\frac{D-26}{6\alpha^{'}}-\frac{1}{2}\nabla^2\phi+\partial^\rho\phi\partial_\rho\phi
-\frac{1}{24}H_{\mu\nu\rho}H^{\mu\nu\rho}
\eea
Considering the condition $\kappa(X^0)\propto\phi^{''}(X^0)$, we take the following ansatz for the graviton
and dilaton backgrounds
\bea\label{ansatz}
\phi^{'}(X^0)&=&\frac{\phi_0}{(X^0)^n},\nn
\kappa(X^0)&=&\frac{\kappa_0}{(X^0)^{n+1}}\nn
\tau_i(X^0)&=&\frac{-\kappa_0}{(X^0)^{n_i}}.
\eea
Note that the constant $\kappa_0$ can be chosen to be the same for $\tau_i$ and $\kappa$ after
a rescaling of the space coordinates $X^i$.
It can easily be seen (c.f. Appendix B) that one needs $n=1$ in order to have homogeneous beta functions.
The solution for the dilaton and the temporal component $g_{00}$ of the metric is then the same as in the case
without antisymmetric-tensor backgrounds~\cite{AEM1}:
\be
\phi=\phi_0\ln(X^0)~~~~~~~~g_{00}=\frac{\kappa_0}{(X^0)^2},
\ee
The non-vanishing components of $H_{\rho\mu\nu}$ are (up to permutations of the indices (0,1,2)):
\be
H_{012}=\varepsilon_{0123}\partial^3 h=-\kappa_0 h_0(X^0)^{(n_3-n_1-n_2-2)/2}.
\ee
Since the direction $X^3$ is singled out, we have $n_1=n_2$, but for the sake of completeness,
we keep these exponents distinct for the moment.
The non-vanishing components of the antisymmetric tensor $B_{\mu\nu}$ are
\be
B_{12}=-B_{21}=-\frac{2\kappa_0 h_0}{n_3-n_1-n_2}(X^0)^{(n_3-n_1-n_2)/2},
\ee
and the non-vanishing one-loop beta functions read (no summation on the index $i$)
\bea
\beta_{00}^{g(1)}&=&\frac{-1}{4(X^0)^2}\left[2h_0^2(X^0)^{n_3}+\sum_{j=1}^3 n_j^2\right] \nn
\beta_{ii}^{g(1)}&=&\frac{n_i}{4(X^0)^{n_i}}\left[2h_0^2(X^0)^{n_3}
+4\phi_0+\sum_{j=1}^3 n_j\right]~~\longrightarrow~~i=1,2 \nn
\beta_{33}^{g(1)}&=&\frac{n_3}{4(X^0)^{n_3}}\left[4\phi_0+\sum_{j=1}^3 n_j\right]\nn
\beta_{12}^{B(1)}&=&\frac{h_0}{4}(X^0)^{(n_3-n_1-n_2)/2}\left[n_3-4\phi_0\right]\nn
\beta^{\phi(1)}&=&-\frac{11}{3\alpha^{'}}+\frac{\phi_0}{4\kappa_0}\left[ 4\phi_0+\sum_{j=1}^3 n_j\right]
-\frac{h_0^2}{4\kappa_0}(X^0)^{n_3}.
\eea
In order for these beta functions to be homogeneous, it is necessary to have $n_3=0$.
The second of the conditions (\ref{conditions}) can be satisfied for late times with $n_1=n_2=2$.
The non-vanishing components of $H_{\rho\mu\nu}$ are then
\be
H_{012}=-\frac{\kappa_0 h_0}{(X^0)^3},
\ee
thereby implying that it is indeed possible, at one-$\sigma$-model-loop, to find homogeneous
beta functions.  We next check that the two-loop order corrections to the Weyl-anomaly coefficients are homogeneous to the one-loop order terms.

\subsection{Second order in $\alpha^{'}$}

In this subsection we check
that the configuration
\bea\label{solution}
g_{\mu\nu}(X^0)&=&\mbox{diag}\left(\frac{\kappa_0}{(X^0)^2},\frac{-\kappa_0}{(X^0)^2},
\frac{-\kappa_0}{(X^0)^2},-\kappa_0\right) \nn
B_{\mu\nu}(X^0)&=&\left(\delta_{\mu 1}\delta_{\nu 2}-\delta_{\mu 2}\delta_{\nu 1}\right)\frac{\kappa_0 h_0}{2(X^0)^2}\nn
\phi(X^0)&=&\phi_0\ln(X^0),
\eea
yields beta functions (Weyl anomaly coefficients) which are homogeneous in $X^0$ at two-$\sigma$-model-loop order. These two-loop expressions
depend on the renormalization scheme that is used~\cite{metsaev}, but their homogeneity does not,
so that it is sufficient to look at the representative terms which appear in the various two-loop beta functions.
These are \cite{metsaev}:
\begin{itemize}
\item for $\beta^g_{00}$ and $\beta^g_{ii}$, $i=1,2$:\\
\bea
R_{0\alpha\beta\gamma}R_0^{~\alpha\beta\gamma}&=&\frac{6}{\kappa_0}(X^0)^{-2}\nn
R_{i\alpha\beta\gamma}R_i^{~\alpha\beta\gamma}&=&-\frac{6}{\kappa_0}(X^0)^{-2}\nn
R^{\alpha\beta\rho\sigma}H_{0\alpha\beta}H_{0\rho\sigma}&=&-\frac{4h_0^2}{\kappa_0}(X^0)^{-2}\nn
R^{\alpha\beta\rho\sigma}H_{i\alpha\beta}H_{i\rho\sigma}&=&\frac{2h_0^2}{\kappa_0}(X^0)^{-2}\nn
H_{\rho\sigma 0}H^{\sigma\alpha\beta}H^\rho_{~\beta\gamma}H^\gamma_{~\alpha 0}
&=&\frac{2h_0^4}{\kappa_0}(X^0)^{-2}\nn
H_{\rho\sigma i}H^{\sigma\alpha\beta}H^\rho_{~\beta\gamma}H^\gamma_{~\alpha i}
&=&-\frac{4h_0^4}{\kappa_0}(X^0)^{-2}\nn
\nabla_0 H_{\alpha\beta\gamma}\nabla_0H^{\alpha\beta\gamma}&=&\frac{54h_0^2}{\kappa_0}(X^0)^{-2}\nn
\eea

\item for $\beta^B_{33}$:
All the contributions vanish.

\item for $\beta^B_{12}$:\\
\bea
R_{1\gamma\alpha\beta}\nabla^\gamma H^{\alpha\beta}_{~~~2}&=&
-\frac{6h_0}{\kappa_0}(X^0)^{-2}\nn
\nabla_\gamma H_{\alpha\beta 1}H_{2\rho}^{~~\alpha}H^{\beta\gamma\rho}&=&-\frac{3h_0^3}{\kappa_0}(X^0)^{-2}\nn
\nabla_\beta \left( H_{\alpha\rho\sigma}H_1^{~~\rho\sigma}\right) H_2^{~~\alpha\beta}&=&\frac{2h_0^3}{\kappa_0}(X^0)^{-2}\nn
H_{\alpha\rho\sigma}H_\beta^{~~\rho\sigma}\nabla^\alpha H^\beta_{~12}&=&-\frac{6h_0^3}{\kappa_0}(X^0)^{-2}\nn
\eea

\item for $\beta^\phi$:\\
\bea
\left( H_{\alpha\beta\gamma}H^{\alpha\beta\gamma}\right)^2&=&\frac{36h_0^4}{\kappa_0^2}\nn
R_{\lambda\mu\nu\rho}R^{\lambda\mu\nu\rho}&=&\frac{6}{\kappa_0^2}\nn
H_{\alpha\beta}^{~~~\mu}H^{\alpha\beta\nu}\nabla_\mu\nabla_\nu\phi&=&\frac{2h_0^2\phi_0}{\kappa_0^2}\nn
R^{\alpha\beta\rho\sigma}H_{\alpha\beta\lambda}H_{\rho\sigma}^{~~~\lambda}&=&-\frac{h_0^2}{\kappa_0^2}\nn
H_{\alpha\beta}^{~~~\mu}H^{\alpha\beta\nu}H_{\gamma\delta\mu}H^{\gamma\delta}_{~~~\nu}&=&\frac{20h_0^4}{\kappa_0^2}\nn
\nabla_\lambda H_{\alpha\beta\gamma}\nabla^\lambda
H^{\alpha\beta\gamma}&=&\frac{54h_0^2}{\kappa_0^2}
\eea

\end{itemize}
As a consequence, the configuration (\ref{solution}) indeed gives beta functions which are homogeneous (in $X^0$)
at two-$\sigma$-model-loop order. We next proceed to discuss the satisfaction of the Weyl invariance conditions in a class of renormalization schemes, which is reached by appropriate redefinitions of the background.

\subsection{String Background redefinitions and Weyl Invariance}

By power counting, it can actually be seen that the configuration (\ref{solution})
leads to homogeneous (in the time $X^0$) beta functions, to all orders in $\alpha^{'}$.
Indeed, whatever power of Ricci or Riemann tensor we consider, multiplied by covariant derivatives of the
dilaton and/or the Kalb-Ramond field strength, contraction of the indices with the metric or its inverse, so as to yield the appropriate tensorial structure for a beta-function term, always leads to
the same power of $X^0$. Thus we have
\bea
\beta^g_{00}&=&\frac{1}{(X^0)^2}\sum_{n=0}^\infty \xi_n(\alpha^{'})^n=\frac{E_0}{(X^0)^2}\nn
\beta^g_{11}&=&\beta^g_{22}=\frac{1}{(X^0)^2}\sum_{n=0}^\infty \zeta_n(\alpha^{'})^n=\frac{E_1}{(X^0)^2}\nn
\beta^g_{33}&=&\sum_{n=0}^\infty \chi_n(\alpha^{'})^n=E_2\nn
\beta^B_{12}&=&\frac{1}{(X^0)^2}\sum_{n=0}^\infty \delta_n(\alpha^{'})^n=\frac{E_3}{(X^0)^2}\nn
\beta^\phi&=&\frac{1}{\alpha^{'}}\sum_{n=0}^\infty \eta_n(\alpha^{'})^n=E_4,
\eea
where the coefficients $\xi_n,\zeta_n,\chi_n,\delta_n,\eta_n$ are $\alpha^{'}$-independent,
and $E_{0,1,2,3,4}$ are constants (we have seen that $E_2$ vanishes at least at two-loops).\\
We discuss now how this homogeneity leads to conformal invariance, using the arguments of \cite{metsaev}.
The latter are perturbative arguments, order-by-order in an $\alpha^{'}$ expansion. It is important to stress at this point that the order-by-order
application of field redefinitions makes them ~{\it local} in target space, which is an essential requirement for the validity of the ``equivalence theorem'', that is the invariance of the (perturbative in $\alpha^{'}$) scattering amplitudes of the string. It is in this sense that the Weyl-invariance of our configuration is guaranteed, namely the existence of a class of renormalization schemes, reached by local field redefinitions, which
lead to a vanishing of the Weyl-anomaly coefficients. \\

We commence our analysis  with the first non-trivial order in $\alpha^{'}$.
At one-$\sigma$-model-loop, the physical predictions of the string happen to be unchanged under the following redefinition of the background fields:
\bea\label{reparametrization}
\tilde g_{\mu\nu}&=&g_{\mu\nu}+\alpha^{'}g_{\mu\nu}\left(a_1 R+a_2 \partial^\rho\phi\partial_\rho\phi
+a_3\nabla^2\phi+a_4H_{\rho\mu\nu}H^{\rho\mu\nu}\right)\nn
\tilde B_{\mu\nu}&=&B_{\mu\nu}+\alpha^{'}\left(b_1\nabla^\rho H_{\rho\mu\nu}+b_2\partial^\rho\phi H_{\rho\mu\nu}\right)\nn
\tilde\phi&=&\phi+\alpha^{'}\left(c_1R+c_2\partial^\rho\phi\partial_\rho\phi+c_3\nabla^2\phi
+c_4H_{\rho\mu\nu}H^{\rho\mu\nu}\right),
\eea
where ($a_1,...,b_1,...,c_1,...$) are free parameters.

It is easy to check that, with our particular configuration (\ref{solution}), the equations (\ref{reparametrization})
do not change the $X^0$ dependence of the fields, since
\bea\label{A123}
\tilde g_{\mu\nu}&=&A_1g_{\mu\nu}\nn
\tilde B_{\mu\nu}&=&A_2B_{\mu\nu}\nn
\tilde\phi&=&\phi+A_3,
\eea
where $A_{1,2,3}$ are constants. The redefinitions (\ref{reparametrization}) change the beta functions,
though, in the following way \cite{metsaev}
\be\label{redefbeta}
\tilde\beta_i=\beta_i+(\tilde g^j-g^j)\frac{\partial\beta^i}{\partial g^j}
-\beta^j\frac{\partial}{\partial g^j}\left(\tilde g^i-g^i\right),
\ee
where $g^i$ is a shorthand notation for $g_{\mu\nu},B_{\mu\nu},\phi$ and  integration over $X^0$ is implied by the repeated-$j$-index notation, in addition to background-species summation.
The set of the ten parameters ($a_1, ...,b_1,...,c_1,...$) is then sufficient to tune the beta functions $\tilde\beta^i$
to zero, thus satisfying Weyl invariance: with the configuration (\ref{solution}), all the terms in the beta functions $\tilde\beta^i$
are again homogeneous functions of $X^0$ (and also homogeneous to the original $\beta^i$),
as a result of the power-law behaviour of the backgrounds with respect to $X^0$:
\be
g_{\mu\nu},~ B_{\mu\nu},~\phi^{''}~\propto ~(X^0)^{-2}.
\ee
Indeed, the terms which appear in $\tilde\beta^g_{00}$, for instance, are typically of the form
\bea
&&\int dX^0\left( \tilde g_{\mu\nu}-g_{\mu\nu}\right)\frac{\delta}{\delta g_{\mu\nu}}\left(\nabla_0\nabla_0\phi\right)\nn
&=&D(A_1-1)\int dX^0\kappa\frac{\delta}{\delta\kappa}\left( \phi^{''}-\frac{\kappa^{'}\phi^{'}}{2\kappa}\right) \nn
&=&\frac{D}{2}(A_1-1)\left( \phi^{''}+\frac{\kappa^{'}}{\kappa}\phi^{'}\right) \nn
&=&-\frac{3}{2}D(A_1-1)\frac{\phi_0}{(X^0)^2},
\eea
which has the expected power law. Similarly,
the terms appearing in $\tilde\beta^\phi$ are typically of the form
\bea
&&\int dX^0\left(\tilde\phi-\phi\right)\frac{\delta}{\delta\phi}\left(\partial^\rho\phi\partial_\rho\phi\right)\nn
&=&A_3\int dX^0\frac{\delta}{\delta\phi}\left(\frac{(\phi^{'})^2}{\kappa}\right)\nn
&=&-2A_3\left(\frac{\phi^{'}}{\kappa}\right)^{'}\nn
&=&-2A_3\frac{\phi_0}{\kappa_0},
\eea
which, as expected, is a constant.
As a consequence, there are only five conditions to satisfy
in order to cancel the beta functions $\tilde\beta^i$:
\bea\label{tildebeta0}
\tilde\beta^g_{00}&=&\frac{\tilde E_0}{(X^0)^2}=0\nn
\tilde\beta^g_{11}&=&\tilde\beta^g_{22}=\frac{\tilde E_1}{(X^0)^2}=0\nn
\tilde\beta^g_{33}&=&\tilde E_2=0\nn
\tilde\beta^B_{12}&=&\frac{\tilde E_3}{(X^0)^2}=0\nn
\nn\tilde\beta^\phi&=&\tilde E_4=0,
\eea
where the constants $\tilde E_{0,1,2,3,4}$ are linear combinations of the parameters ($a_1, ...,b_1,...,c_1,...$).
It is then possible to cancel the two-loop beta functions, and thus guarantee Weyl invariance to this order, by
implementing the redefinition (\ref{reparametrization}), independently of the amplitudes $\phi_0,h_0$.
A given solution of the conditions (\ref{tildebeta0})
in terms of the parameters ($a_1, ...,b_1,...,c_1,...$)
leads to a specific set of the constants $A_{1,2,3}$ defined in eqs.(\ref{A123}),
but leaves the $X^0$ dependence of the background fields unchanged.
The next step is to argue, by induction, that similar results hold to any order in $\alpha^{'}$. This has been
explained in detail in ref.~\cite{AEM1} for graviton and dilaton backgrounds only, but can be extended straightforwardly to our case with a non-trivial Kalb-Ramond field. As a consequence, we conjecture that the configuration (\ref{solution})
satisfies Weyl invariance in a non-perturbative way, to all orders in $\alpha^{'}$,
for any amplitudes $\phi_0,h_0$.

\section{Cosmological properties of the $\lambda$-flow-fixed-point solution}

In this section we shall discuss the physical (cosmological) properties of our non-trivial $\lambda$-flow-fixed-point solution (\ref{solution}), and demonstrate that it leads to an anisotropically expanding Robertson-Walker Universe.
To this end we pass onto the Einstein-frame metric, in which the Einstein curvature term (lowest order in an $\alpha^{'}$-expansion) in the target-space effective action is canonically normalized.
This is viewed as the ``physical metric'', where cosmological measurements are performed.
The relation between the physical metric in the Einstein frame and the string ($\sigma$-model)-frame metric is given in  $D=4$ space-time dimensions
by~\cite{ABEN2},
\bea\label{einsteinframe}
ds^2&=&dt^2-a_1^2(t)(dx^1)^2-a_2^2(t)(dx^2)^2-a_3^2(t)(dx^3)^2\nn
&=&\exp\left\{-2\phi(x^0)\right\}g_{\mu\nu}dx^\mu dx^\nu\nn
&=&\exp\left\{-2\phi_0\ln(x^0)\right\}\kappa_0\left[\left(\frac{dx^0}{x^0}\right)^2
-\left(\frac{dx^1}{x^0}\right)^2-\left(\frac{dx^2}{x^0}\right)^2-(dx^3)^2\right],
\eea
where the $x^\mu$ denote the zero modes of $X^\mu$, and $a_i(t)$ are the scale factors along the
spatial directions $i=1,2,3$. We then have:
\be
\frac{dt}{dx^0}=\pm\sqrt{\kappa_0} (x^0)^{-1-\phi_0},
\ee
which implies that the Einstein-frame cosmic time $t$ is:
\be\label{tx^0}
t=\frac{\sqrt{\kappa_0}}{|\phi_0|}(x^0)^{-\phi_0}.
\ee
Finally, the components of the scale factor read in this frame
\bea\label{scalefactors}
a_1(t)=a_2(t)&=&a_0~t^{1+1/\phi_0}\nn
a_3(t)&=&a_0~t
\eea
where $a_0$ is a constant. As a consequence, the effect of the antisymmetric tensor field, in the configuration (\ref{solution}), is to generate a
{\it linear expansion} in the direction which is singled out by the axion field. Note that, if $\phi_0=-1$, the directions $X^1,X^2$ are static, as was the case
without antisymmetric tensor \cite{AEM1}. It remains to be seen whether there are other, more general, solutions of the $H_{\mu\nu\rho}$ field, which could lead to different types of Universes. Such issues will constitute the subject of a forthcoming publication.

It can be mentioned in passing at this point that the above features can be
carried out in a qualitatively similar manner to a higher-dimensional case.
In fact, the property that the dimension singled out by the antisymmetric tensor field increases linearly with the target time, can be used to discuss cosmological brane models
with large bulk dimensions, starting from relatively small ones in early stages of the Universe.

There is an issue with the eventual exit of the Universe (\ref{scalefactors}) from its linearly expanding  anisotropic phase. As demonstrated in \cite{AEM1}, in the absence of the Kalb-Ramond field, upon the choice of the dilaton amplitude $\phi_0 = -1$, one obtains a four-dimensional Minkowski space time, as an asymptotic equilibrium point. This probably indicates that, in order to secure an exit from the phase described by the metric configuration (\ref{scalefactors}), one needs  time-dependent configurations for the axion field, which relax to zero asymptotically in cosmic time. Whether such configurations constitute appropriate Weyl-invariant fixed points of the $\lambda$-flow, or can be treated in a more general setting, by going away from such fixed points, remains to be seen. We hope to come back to such issues in a forthcoming work.

We next remark that, as is easy to see, in the Einstein frame, the dilaton is independent of the amplitude $\phi_0$ and reads,
up to a constant,
\be\label{dilaton}
\phi(t)=-\ln t.
\ee

The presence of a time dependent dilaton imposes important constraints on the various physical couplings that characterize our effective low-energy field theory derived from strings. In particular, they will imply time-dependent
gauge couplings, and thus variable ``fine structure constant'', which could be constrained by experiment. Moreover, the specific coupling of the antisymmetric tensor with the electromagnetic field, which is a characteristic feature of our string-inspired model, will lead to non-trivial optical activity for the (anisotropic) Universe (\ref{scalefactors}).
In the next section we discuss briefly such phenomenological consequences of our solution, with the aim of setting the range for its phenomenological validity.

\section{Coupling the Antisymmetric Tensor to Electromagnetic fields and Non-trivial Optical Activity}

We consider here $\phi_0=-1$, in which case, as discussed previously,  we have static $x^1,x^2$ directions and a linearly expanding $x^3$
direction. We can also take $x^0=t$ after rescaling the time by $\sqrt{\kappa_0}/|\phi_0|$ (see eq.(\ref{tx^0})),
and in what follows we keep $\alpha^{'}=1$, so that the non-perturbative background-field configuration is:
\bea\label{configg}
g_{\mu\nu}(t)&=&\mbox{diag}(1,-1,-1,-t^2)\nn
H_{012}(t)&=&-\frac{\kappa_0 h_0}{t^3}\nn
\phi(t)&=&-\ln t.
\eea
In order to take into account the anisotropy in the electromagnetic propagation over cosmological scales~\cite{nodland},
we couple the massless excitations of the string to a $U(1)$ gauge field by the
introduction of the following modified three-form field strength $\tilde H_{\rho\mu\nu}$:
\be\label{atmod}
\tilde H_{\rho\mu\nu}=H_{\rho\mu\nu}+\frac{1}{M}A_{[\rho}F_{\mu\nu]},
\ee
where $M$ is a scale with dimensions of mass, which depends on the details of the underlying microscopic string/brane model; if the original theory lives in its critical dimensionality, which is higher than four, then
the corresponding Chern Simons term in (\ref{atmod})
arises from the compactification of the higher dimensions,
in which case $M$ depends, in addition to the string mass scale $M_s$, also
on the size of the extra dimensions, and in general the details of the compactification procedure~\cite{coupling,birefringence}. In our $D=4$ solution, $M$ can be identified directly the string mass scale, $M_s$:  $M = M_s = 1/\ell_s = (\alpha^{'})^{-1/2}$, since the latter is the only dimensionful free parameter in the model.

In order to recover gauge invariance,
one should remember that a gauge transformation on $A_\mu$ has to be followed by a gauge transformation on the
antisymmetric tensor $B_{\mu\nu}$. Indeed, since $\partial_{[\rho}F_{\mu\nu]}=0$,
the field strength $\tilde H_{\rho\mu\nu}$ is invariant under the simultaneous transformations
\bea
A_\mu&\to& A_\mu+\partial_\mu\psi\\
B_{\mu\nu}&\to& B_{\mu\nu}-\frac{\psi}{M} F_{\mu\nu}\nonumber.
\eea
We are interested in the coupling of the electromagnetic field to the non-perturbative background (\ref{configg}).

A complete treatment, leading to an all-orders-in-$\alpha^{'}$ solution
for all background fields, including the electromagnetic, would require
considering the exact expression for the contributions of the electromagnetic field to the graviton, dilaton, and antisymmetric-tensor beta functions, as well as its own beta function in the presence of the curved gravitational background. This is not known at present in a general string theory model.
In open string theories, such as the ones pertaining to brane models,
the gauge field dynamics is partly
described by the $\alpha^{'}$-non-perturbative
Born-Infeld Lagrangian~\cite{low.ener.eff.}, but in a general string theory  the contributions mixing the electromagnetic gauge field strength to the gravitational curvature parts of the effective action, are not known in a closed form to all orders in $\alpha^{'}$.

Fortunately, for our phenomenological purposes in this section, where we are interested in the effects of the electromagnetic field in late epochs of the Universe, a lowest-non-trivial-order in $\alpha^{'}$ analysis suffices.
What we shall do, therefore, is to consider the solution of the electromagnetic field equations of motion in the background of (\ref{configg}) of the gravitational multiplet of the string, without considering back reaction effects of the electromagnetic potential onto the space time. This procedure was also applied previously in the literature~\cite{birefringence} in order to study similar effects in string theory.
Nevertheless, at the end of the section we shall present qualitative arguments as to why the conclusions from the analysis in this section can be carried out qualitatively to the full problem, where the full Weyl-invariant
$\alpha^{'}$-non-perturbative solution, including back reaction effects of the gauge potential onto the gravitational multiple of the string is considered. As we shall discuss below, this can be achieved due to the specific time ($x^0$) dependence of the lowest-order solution of the electromagnetic field, which does not affect qualitatively the homogeneity of the beta functions.

Let us commence our analysis by
considering the effective action for the massless degrees of freedom
corresponding to one-$\sigma$-model-loop Weyl invariance, which is, in the Einatein frame \cite{low.ener.eff.}:
\be\label{loweneffact}
S_{eff}=\int d^4x \sqrt{-g}\left\lbrace R-2\partial_\mu\phi\partial^\mu\phi
-\frac{e^{-4\phi}}{12}\tilde H_{\rho\mu\nu}\tilde H^{\rho\mu\nu}
-\frac{e^{-2\phi}}{4}F_{\mu\nu}F^{\mu\nu}\right\rbrace.
\ee
The reader should recall, at this point, that to this order in $\alpha^{'}$, this bosonic part of the string effective action has a universal structure among all known string theories.
Plugging the configuration (\ref{configg}) in this low energy effective action, we find that
the relevant effective action for the gauge field is
\be\label{gauge}
S_{gauge}=\int dt d\vec x~t\left\lbrace -\frac{t^2}{4}F_{\mu\nu}F^{\mu\nu}
+\epsilon t A_{[0}F_{12]}\right\rbrace,
\ee
where
\be\label{deffeps}
\epsilon=\frac{\kappa_0h_0}{M}.
\ee
One should keep in mind that the action (\ref{gauge}) is valid in a given gauge, since it does not
contain the terms in $B_{\mu\nu}$ which would compensate a gauge transformation on $A_\mu$.
This action exhibits a time-dependent fine structure constant $\alpha$ satisfying~\footnote{Caution should be exercised at this point. To properly account for the time-dependence of the
gauge couplings, one should consider the fermionic (e.g. electron ) part of the matter action, with canonically normalized kinetic terms, after appropriate rescalings with exponential dilaton factors. In some models~\cite{damourpolyakov} the Einstein-frame action (\ref{loweneffact}) satisfies the above-mentioned normalization criteria, and hence the (experimentally measurable) running fine structure constant in such cases is
given by (\ref{rcc}).}:
\be\label{rcc}
\alpha^{-1}\propto e^{-2\phi}=t^2,
\ee
such that
\be\label{ourresult}
\frac{|\Delta\alpha|}{\alpha}=2\frac{|\Delta t|}{t},
\ee
which can be compared to data. Indeed, after a long and not always uncontroversial series of experimental
measurements, the consensus seems to be that currently there is no experimental evidence for
a time-varying fine structure constant.
Recent measurements~\cite{alpha} (2004), using High-redshift quasars absorption-line Mg II systems (in the redshift range $0.4 < z < 2.3$,
with the median redshift of the sample $z \sim 1.55$,
corresponding to a look-back time of $\Delta t \sim 9 Gyr$ in the most favoured standard cosmological model) have given the constraint
$\frac{\Delta \alpha}{\alpha} = \left(-0.06 \pm 0.06\right)\times 10^{-5}~.$
If one assumes that the age of the Universe is $t \sim 14 ~{\rm Gyr}$, then the result (\ref{ourresult}) leads to the estimate
$\Delta \alpha / \alpha \sim 1.3 $, which is too large to represent reality.

However, before dismissing the model on such phenomenological grounds, one should take into account that the solution (\ref{configg}) represents~\cite{AEM1} an asymptotic (in cosmic time) situation~\footnote{This follows from the fact that
the solution is a non-trivial fixed point of the $\lambda$-flow, which from a Wilsonian renormalization-group point of view corresponds~\cite{AEM1} to a world-sheet infrared fixed point. The latter is customarily considered as an asymptotic equilibrium point of the system under study, in this case the Universe. Such an analysis has been made in \cite{AEM1} for graviton and dilaton backgrounds only, but can be adapted
here in a rather straightforward manner.} and not necessarily a present-day epoch of the Universe. The complete evolution of the Universe in our context, including the current era, probably necessitates
the solution of the $\lambda$-flow equations away from the fixed points, which so far has not been studied properly. Only when such an analysis is made can one perform a proper study of the running-with-cosmic-time gauge couplings, and compare them with phenomenological data.
This is left for a future work.

We next study the implications of the effective action (\ref{gauge}) for cosmic optics. Again, the reader should bear in mind that, in view of
the phenomenologically non-realistic metric configuration (\ref{configg}),
our findings are only indicative and do not represent a situation that could be compared with observations.
The equations of motion for the electromagnetic field in the background (\ref{configg}) read:
\bea\label{newMax}
{\bf\nabla}\cdot{\bf\tilde B}&=&0\\
{\bf\nabla}\cdot{\bf E}&=&\frac{2\epsilon}{t}B_3\nn
{\bf\nabla}\times{\bf\tilde E}&=&-\partial_t{\bf\tilde B}\nn
\left({\bf\nabla}\times{\bf B}\right)_1&=&\partial_t E_1+\frac{3}{t}E_1
-\frac{2\epsilon}{t} E_2+\frac{2\epsilon}{t^2}A_2,\nn
\left({\bf\nabla}\times{\bf B}\right)_2&=&\partial_t E_2+\frac{3}{t}E_2+
\frac{2\epsilon}{t} E_1-\frac{2\epsilon}{t^2}A_1,\nn
\left({\bf\nabla}\times{\bf B}\right)_3&=&\partial_t E_3+\frac{5}{t}E_3\nonumber
\eea
where we have defined:
\bea
&&{\bf E}=(E_1,E_2,E_3),~~~~~~{\bf B}=(B_1,B_2,B_3)\nn
&&{\bf\tilde E}=(E_1,E_2,t^2E_3),~~~~{\bf\tilde B}=(t^2B_1,t^2B_2,B_3),
\eea
and $E_i,B_j$ are the usual components of the electric and magnetic fields.
The appearance of the gauge components $A_{1,2}$ in the generalized Maxwell's equations (\ref{newMax})
is a consequence of the gauge-fixed action (\ref{gauge}).
In the case of a plane wave propagating in the $x^3$ direction,
the time evolution of the electric field is then given by the following equation,
where higher powers of $1/t$ have been ignored,
\be\label{evolE}
{\bf\ddot E}+\frac{3}{t}{\bf\dot E}+\frac{2\epsilon}{t}{\bf \dot E_\bot}=0,
\ee
where ${\bf E}=(E_1,E_2,0)$, ${\bf E_\bot}=(-E_2,E_1,0)$ and the overdot represents time derivative.
Note that the spatial-derivative term $\partial^3\partial_3$, which should appear in a wave equation, is of order $1/t^2$ and
is therefore not taken into account here.
The solution of the equation (\ref{evolE}) is, up to an additive constant,
and using complex notations in the plane $(x^1,x^2)$,
\be\label{solE}
{\bf E}=\frac{{\bf K_0}f(x^3)}{t^2}\exp\left(-2i\epsilon\ln t\right),
\ee
where ${\bf K_0}$ is a complex constant and $f$ is a function of the dimensionless space coordinate $x^3$.
From the equations of motion (\ref{newMax}), then, it follows that the magnetic field satisfies the equation:
\be
{\bf\nabla}\times{\bf B}=\frac{{\bf K_0}f(x^3)}{t^3}\exp\left(-2i\epsilon\ln t\right),
\ee
It is possible, then, to consider a situation where ${\bf E}$ and ${\bf B}$ are perpendicular to each other and
lie both on the plane $(x^1,x^2)$.
If we take for the spatial dependence $f(x^3)=\cos(x^3)$, the electromagnetic field is, in the plane $(x^1,x^2)$,
\bea\label{solEB}
{\bf E}&=&{\bf K_0}\exp\left(-2i\epsilon\ln t\right)\frac{\cos(x^3)}{t^2}\nn
{\bf B}&=&i{\bf K_0}\exp\left(-2i\epsilon\ln t\right)\frac{\sin(x^3)}{t^3}.
\eea
We can now check the consistency of this solution as
far as conformal invariance is concerned. From the expression (\ref{solEB}) for the electromagnetic field  we obtain:
\be\label{F^2}
F_{\mu\nu}F^{\mu\nu}=2\left(|{\bf E}\cdot{\bf\tilde E}|+|{\bf B}\cdot{\bf \tilde B}|\right)=2\frac{|{\bf K_0}|^2}{t^4},
\ee
and therefore the corresponding term in the effective action (\ref{loweneffact}) is of the same order as the others:
\be
e^{-2\phi}F_{\mu\nu}F^{\mu\nu}\propto e^{-4\phi}H_{\mu\nu\rho}H^{\mu\nu\rho}\propto \partial_\mu\phi\partial^\mu\phi
\propto t^{-2}.
\ee
This homogeneity can also be seen at the level of the beta function $\beta^g_{\mu\nu}$ as follows:
according to the general discussion in stringy $\sigma$-models presented in \cite{osborn}, the non-vanishing beta functions are proportional to off-shell variations of the target-space effective action with respect to the background fields $g^i$:
\be
 \frac{\delta S_{eff}}{\delta g^i} = {\cal G}_{ij} \beta^j
 \label{offshell}
 \ee
 where ${\cal G}_{ij}$ is related to the the world-sheet two-point correlation functions of vertex
operators $<V_i (z) V_j(0) >$ in the way explained in \cite{osborn}, and is positive definite, invertible,
playing the r\^ole of a (Zamolodchikov) metric in string theory space $\{ g^i \}$.
From a target-space viewpoint, this metric is an infinite series in $\alpha^{'}$, and involves in general derivative terms acting on the beta functions.

To lowest order in $\alpha^{'}$, which suffices for our phenomenological purposes in this section, the Zamolodchikov metric leads to terms quadratic in the metric tensor~\cite{MM}. Hence, the off-shell
variation of the effective action (\ref{loweneffact}) with respect to the metric
leads to the following expression, linear in the various beta functions~\cite{MM}:
\be
\frac{e^{2\phi}}{\sqrt{-g}}\frac{\delta S_{eff}}{\delta g^{\mu\nu}}=
g_{\mu\nu}\left(g^{\alpha\beta}\beta^g_{\alpha\beta}+\beta^\phi\right)+\cdot\cdot\cdot,
\ee
where the dots represent other operators, given in \cite{MM}, which are all homogeneous to the ones given here.
In the string frame, the above expression is of order $(x^0)^{-2}$. We now check that this $x^0$-dependence
is indeed the one obtained from the gauge field contribution. The latter is given by:
\be\label{prout}
\frac{e^{2\phi}}{\sqrt{-g}}\frac{\delta}{\delta g^{\mu\nu}}\left( -\frac{1}{4}
\int d^4x\sqrt{-g} ~e^{-2\phi}F^2\right)=
\frac{1}{8}g_{\mu\nu}F^2-\frac{1}{2}F_{\rho\mu}F_{\sigma\nu}g^{\rho\sigma}.
\ee
In order to find the corresponding power in $x^0$, we need to compute $F^2$ in the string frame,
and use the fact that, for the solution (\ref{solEB}), we have $F_{0i}\propto t^{-2}\propto (x^0)^{-2}$
and $F_{i3}\propto t^{-1}\propto (x^0)^{-1}$.
As a consequence, $F^{0i}\propto (x^0)^2$, $F^{i3}\propto (x^0)^3$, since $g^{\alpha\beta}\propto (x^0)^2$.
Therefore $F^2\propto (x^0)^0$=constant,
and the corresponding functional derivative (\ref{prout}) is indeed of the expected power $(x^0)^{-2}$.
This homogeneity, then, implies that one may be able of finding an appropriate background-field redefinition to achieve the vanishing of the pertinent beta functions in some renormalization scheme, thereby ensuring Weyl invariance of our space-time solution in the presence of (back reaction effects of) the electromagnetic field, to lowest non-trivial order in $\alpha^{'}$.

The above conclusion can be carried through to higher orders in $\alpha^{'}$. To see this, consider the simple case of the Abelian Born-Infeld (BI) Lagrangian~\cite{born}, which is argued to characterize the Abelian gauge field dynamics of open string models, encountered, for instance, in D-brane theories. The BI Lagrangian involves a re-summation to all orders in $\alpha^{'}$~\cite{born}. In a string-frame background metric $G_{\mu\nu}$  and dilaton $\phi$ the tree-level BI Lagrangian reads:
\be
{\cal L}_{BI}\propto e^{-2\phi}\left\{
\left(-{\rm det}[G_{\mu\nu}]\right)^{1/2} - \left(-{\rm det}[ G_{\mu\nu} + 2\pi \alpha^{'}F_{\mu\nu} ]\right)^{1/2}
\right\}
\label{biaction}
\ee
In four dimensional D=4 space time, we have the identity~\cite{born}:
\be\label{four}
{\rm det}_4\left[ G_{\mu\nu} + 2\pi \alpha^{'}F_{\mu\nu} \right]  =
{\rm det}_4[G_{\mu\nu}] \left( 1 + \frac{I_1}{2b}-\frac{I_2^2}{16b^2}\right),
\ee
where $b^{-1} \equiv (2\pi \alpha^{'})^2$ and
\bea
I_1 &=& F_{\mu\nu}F^{\mu\nu}\nn
I_2 &=& F_{\mu\nu}~^{\star}F^{\mu\nu}~~~~~~\mbox{with}~~~~~~
^{\star}F^{\mu\nu} \equiv \frac{1}{2}\epsilon^{\mu\nu\rho\sigma}F_{\rho\sigma}.
\eea
In the Einstein-frame, $g_{\mu\nu}=e^{-2\phi}G_{\mu\nu}$
and the relevant BI part of the tree-level string effective action reads then \cite{bidil}:
\be
{\cal L}_{BI}\propto
\sqrt{-g} ~e^{2\phi}\left\{ 1-\sqrt{1 + \frac{I_1^{(E)}}{2b}-\frac{(I_2^{(E)})^2}{16b^2}}\right\},
\label{dilbiaction}
\ee
where the subscript $^{(E)}$ denotes the Einstein frame.
From the previous discussion on the $x^0$-dependence of the components of the gauge field strength,
the BI part of the string effective Lagrangian scales therefore like the gravitational part, i.e. $\propto t^{-2}$.

The situation gets formally more complicated when we consider the dynamics of the gravitational field (Riemann curvature tensors {\it etc.}), obtained by adding (\ref{dilbiaction}) to the non-gauge part of the action (\ref{loweneffact}) and taking into account the correction terms of higher order in $\alpha^{'}$. The precise structure of such terms in the presence of the gauge fields, beyond the first few orders in $\alpha^{'}$,  is not known in  a closed form in general.
Nevertheless, by checking explicitly the few low-orders in $\alpha^{'}$ for which there are available expressions in the literature~\cite{ross}, we can verify the homogeneity of those higher order mixed gauge-gravitational terms in the beta functions as compared with the lowest-order ones.
In this way we are able to argue in favour of/conjecture the Weyl invariance of our non-perturbative solution (\ref{configg}), obtained in section 3, in the presence of gauge fields, to all orders in $\alpha^{'}$.

We now turn to the phenomenological implications of the solution (\ref{solEB}) as far as the non-trivial optical activity is concerned.
As can be seen from this solution,
the friction term proportional to ${\bf\dot E}$ in eq.(\ref{evolE}), which is due to the dilaton,
implies a damping of the (amplitude of the) electromagnetic field strength with the cosmic time. Moreover, the antisymmetric tensor three-form
generates a non-trivial optical activity, arising from the term proportional to ${\bf \dot E_\bot}$ in  eq.(\ref{evolE}).
The corresponding optical rotation angle is obtained by the phase difference
\be
\Delta(t)=|\arg({\bf E})-\arg({\bf E}^*)|=4\epsilon\ln t = 4\frac{\kappa_0 h_0}{M}\ln t,
\ee
and is therefore logarithmic in cosmic time.

Thus, spatially anisotropic expansion and non-trivial optical activity in the
(string) Universe could be of common origin, namely that of a non-trivial Kalb-Ramond antisymmetric tensor field of the string multiplet, playing the r\^ole of a four-dimensional axion pseudoscalar field. As already stressed, of course,
it remains to be seen
whether phenomenologically realistic cosmological models could be constructed in this non-perturbative (in $\alpha^{'}$) framework, in the sense of providing a complete evolution of the Universe. As already mentioned, this may probably necessitate solving the $\lambda$-flow equations away from the fixed points, while keeping the Weyl invariance. We hope to study such issues in the near future.

\section{Conclusions and Outlook}

In this work we have applied a previously-developed novel renormalization-group approach to stringy $\sigma$-models in graviton, dilaton and Kalb-Ramond antisymmetric tensor backgrounds, in order to construct four-dimensional non-perturbative (in $\alpha^{'}$) solutions to the Weyl invariance conditions, of cosmological interest.

We found that, upon assuming a specific ansatz for the antisymmetric tensor field, corresponding to a non-trivial four-dimensional axion field singling out a spatial direction, one obtains an anisotropic Universe, which is linearly expanding along the direction singled out by the axion, and static in the others. This result could also be extended to higher-dimensional cases, to serve as a model for obtaining large bulk dimensions (expanding with the cosmic time), as a result of specific configurations of appropriate tensor gauge fields, with non-trivial fluxes along those directions.

The most important phenomenological consequences of our approach come from the coupling of the antisymmetric tensor to the electromagnetic field, in a way appropriate to string theory. It must be noted at this point that, unlike
the case where the electromagnetic field was absent, discussed in section 3,
in its presence we
are not in a position to prove rigorously, but only argue about, the Weyl invariance of our solution to all orders in $\alpha^{'}$. This is due to the fact that
the various terms
in the beta functions that couple the gauge and field dynamics are not known in a closed form in general. Nevertheless, in our work we have been able to demonstrate the Weyl invariance of our solution in the presence of Abelian gauge fields explicitly to lowest order in $\alpha^{'}$ , where the beta functions are explicitly known and have a universal form in all string theories.  Such a truncation was sufficient for our phenomenological purposes, of demonstrating non trivial optical activity of our solution at late eras of the string Universe, characterized by weak curvature and gauge fields.

In this respect, we also mention that our analysis in this work and in \cite{AEM1} was restricted to string tree-level effective actions. At late eras of the Universe, the specific dilaton configuration (\ref{configg}) implies a weak string coupling $g_s ={\rm exp(\phi)}$, thereby making the contributions from higher world-sheet topologies (string loops) negligible. On the other hand, at early epochs of the Universe, string-loop effects may become important, leading to form factors in front of the various terms in the effective action~\cite{stringcosmo}, which re-sum different powers of $g_s(\phi)$. We have not analyzed such situations in our context. This is something, however, that eventually should be done, in order to be able of understanding strong string coupling situations within the approach.

As the main phenomenological aspects  of our results in this article, we mention the time-dependent fine structure constant and a non-trivial optical activity, namely the rotation of the direction of the electric field with the cosmic time, in a way  depending on the details of the axion configuration. Our findings, however, are only indicative, given that the resulting space time is not a phenomenologically realistic one that could describe the present era of our Universe. Thus, the predictions of our model cannot be compared to real astrophysical data.

To this end, one should really reconsider the analysis in realistic Robertson-Walker space time backgrounds, which describe various eras of the Universe. Whether such space times constitute fixed points of the $\lambda$-flow, as demonstrated for the configuration (\ref{configg}), is not yet known. This is an issue to be discussed in a forthcoming publication.
Only when such an analysis becomes available, can one make realistic predictions that could be compared with observations. Nevertheless, we believe that the results of the current work are sufficiently interesting to encourage further studies in string systems along the above directions.

\section*{Acknowledgments}

The work of N.E.M. is partially supported by the European Union through the FP6 Marie Curie Research and Training Network \emph{UniverseNet} (MRTN-CT-2006-035863).

\section*{Appendix A: Derivation of the evolution for the effective action}

In this Appendix we outline the derivation of the flow with respect to the control parameter of the quantum fluctuations in a bosonic stringy $\sigma$-model with graviton, dilaton and antisymmetric tensor backgrounds.
We start with the bare action for the (first quantized) bosonic string on a Euclidean world sheet, expressed in terms of
microscopic fields $\tilde X^\mu$ defined in the bare theory:
\be
S=\frac{1}{4\pi}\int d^2\xi\sqrt{\gamma}\left\{\lambda\left(
\gamma^{ab}\eta_{\mu\nu}+\varepsilon^{ab}a_{\mu\nu}\right) \partial_a\tilde X^\mu\partial_b\tilde X^\nu
+R^{(2)}\phi_{bare}(\tilde X^0)\right\},
\ee
to which we add the source term
\be
S_S=\int d^2\xi\sqrt{\gamma}R^{(2)}\eta_{\mu\nu}V^\mu\tilde X^\nu,
\ee
in order to define the classical fields. The corresponding partition function $Z$ and
the generating functional of the connected graphs $W$ are
related as usual:
\be
Z=\int{\cal D}[\tilde X]e^{-S-S_S}=e^{-W}.
\ee
The classical fields $X^\mu$ are defined by
\be
X^\mu=\frac{1}{Z}\int{\cal D}[\tilde X]\tilde X^\mu e^{-S-S_S}=\frac{1}{Z}\left<\tilde X^\mu\right>,
\ee
and are obtained by differentiating $W$ with respect to the source $V_\mu$:
\be\label{VX}
\frac{1}{\sqrt{\gamma_\xi}R^{(2)}_\xi}\frac{\delta W}{\delta V_\mu(\xi)}=X^\mu(\xi).
\ee
The second derivative of $W$ is then
\be
\frac{1}{\sqrt{\gamma_\zeta\gamma_\xi}R^{(2)}_\zeta R^{(2)}_\xi}\frac{\delta^2 W}{\delta V_\mu(\zeta)\delta V_\nu(\xi)}
=X^\nu(\xi)X^\mu(\zeta)-\frac{\left<\tilde X^\nu(\xi)\tilde X^\mu(\zeta)\right>}{Z}.
\ee
Inverting the relation (\ref{VX}) between $V_\mu$ and $X^\mu$,
we then introduce the Legendre transform of $W$, namely
the functional $\Gamma$ responsible for the generation of
proper graphs:
\be
\Gamma=W-\int d^2\xi\sqrt{\gamma}R^{(2)}V^\mu X_\mu.
\ee
The functional derivatives of $\Gamma$ are:
\bea\label{derivG}
\frac{1}{\sqrt{\gamma_\xi}R^{(2)}_\xi}\frac{\delta\Gamma}{\delta X^\mu(\xi)}&=&-V_\mu(\xi),\nn
\frac{1}{\sqrt{\gamma_\xi\gamma_\zeta}R^{(2)}_\xi R^{(2)}_\zeta}
\frac{\delta^2\Gamma}{\delta X^\nu(\zeta)\delta X^\mu(\xi)}&=&
-\left(\frac{\delta^2W}{\delta V_\nu(\zeta)\delta V_\mu(\xi)}\right)^{-1}.
\eea

\vspace{.5cm}

\nin The evolution of $W$ with $\lambda$ is given by
\bea
\dot W
&=&\frac{1}{4\pi Z}\int d^2\xi\sqrt{\gamma}\left(\gamma^{ab}\eta_{\mu\nu}+\varepsilon^{ab}a_{\mu\nu}\right)
\left<\partial_a\tilde X^\mu\partial_b\tilde X^\nu\right>\\
&=&\frac{1}{4\pi Z}\mbox{Tr}\left\{\left( \gamma^{ab}\eta_{\mu\nu}+\varepsilon^{ab}a_{\mu\nu}\right)
\frac{\partial}{\partial\xi^a}\frac{\partial}{\partial\zeta^b}
\left<\tilde X^\mu(\xi)\tilde X^\nu(\zeta)\right>\right\},\nonumber
\eea
where the overdot denotes derivative with respect to $\lambda$, and the trace is
defined as:
\be\label{trace}
\mbox{Tr}\{\cdot\cdot\cdot\}=
\int d^2\xi d^2\zeta\sqrt{\gamma_\xi\gamma_\zeta}\{\cdot\cdot\cdot\}\delta^2(\xi-\zeta).
\ee
We then obtain
\bea\label{evolW}
\dot W&=&\frac{1}{4\pi}\int d^2\xi\sqrt{\gamma}\left(\gamma^{ab}\eta_{\mu\nu}+\varepsilon^{ab}a_{\mu\nu}\right)
\partial_a X^\mu\partial_b X^\nu\nn
&&-\frac{1}{4\pi}\mbox{Tr}\left\{\left(\gamma^{ab}\eta_{\mu\nu}+\varepsilon^{ab}a_{\mu\nu}\right)
\frac{\partial}{\partial\xi^a}\frac{\partial}{\partial\zeta^b}
\left(\frac{1}{\sqrt{\gamma_\xi\gamma_\zeta}R^{(2)}_\xi R^{(2)}_\zeta}
\frac{\delta^2 W}{\delta V_\mu(\zeta)\delta V_\nu(\xi)}\right)\right\}\nonumber.
\eea
Finally, the evolution of $\Gamma$ is obtained by noting
that the independent variables of
$\Gamma$ are $X^\mu$ and $\lambda$, and that
\bea\label{dotG}
\dot\Gamma&=&\dot W+\int d^2\xi\frac{\partial W}{\partial V_\mu}
\dot V_\mu-\int d^2\xi\sqrt\gamma R\dot V_\mu X^\mu\nn
&=&\dot W.
\eea
Using eqs.(\ref{derivG}),
(\ref{evolW}) and (\ref{dotG}), then,
we obtain the following evolution equation
for $\Gamma$:
\bea\label{evolGappendix}
\dot\Gamma&=&\frac{1}{4\pi}\int d^2\xi\sqrt{\gamma}\left(\gamma^{ab}\eta_{\mu\nu}+\varepsilon^{ab}a_{\mu\nu}\right)
\partial_a X^\mu\partial_b X^\nu\nn
&&+\frac{1}{4\pi}\mbox{Tr}\left\{\left(\gamma^{ab}\eta_{\mu\nu}+\varepsilon^{ab}a_{\mu\nu}\right)
\frac{\partial}{\partial\xi^a}\frac{\partial}{\partial\zeta^b}
\left(\frac{\delta^2\Gamma}{\delta X^\mu(\zeta)\delta X^\nu(\xi)}\right)^{-1}\right\}.\nonumber
\eea

\noindent
We now assume the following functional dependence:
\bea
\Gamma&=&\frac{1}{4\pi}\int d^2\xi\sqrt{\gamma}\left\{\gamma^{ab}\kappa_\lambda(X^0)\partial_a X^0\partial_b X^0
+\gamma^{ab}\sum_{i=1}^{D-1}\tau_i(X^0)\partial_a X^i\partial_b X^i\right.\nn
&&~~~~~~~~~~~~~~~+\varepsilon^{ab}B_{\mu\nu}(X^0)\partial_a X^\mu\partial_b X^\nu+R^{(2)}\phi_\lambda(X^0)\Bigg\}.
\eea
With the metric $\gamma^{ab}=\delta^{ab}$ and constant configurations $X^\mu(\xi)=x^\mu$,
the second functional derivatives of $\Gamma$ are (no summation on $i$)
\bea
\frac{\delta^2\Gamma}{\delta X^0(\zeta)\delta X^0(\xi)}&=&
-\frac{\kappa}{2\pi}\Delta\delta^2(\xi-\zeta)+\frac{R^{(2)}\phi^{''}}{4\pi}\delta^2(\xi-\zeta),\nn
\frac{\delta^2\Gamma}{\delta X^i(\zeta)\delta X^i(\xi)}&=&
-\frac{\tau_i}{2\pi}\Delta\delta^2(\xi-\zeta)\nn
\frac{\delta^2\Gamma}{\delta X^i(\zeta)\delta X^j(\xi)}&=&0~~~~~~~i\ne j
\eea
where a prime denotes derivative with respect to $x_0$. Note that, for constant configurations $X^\mu(\xi)=x^\mu$,
the antisymmetric tensor $B_{\mu\nu}$ does not play a role in the evolution equation for $\Gamma_\lambda$, as
its contribution is proportional to
\be
\varepsilon^{ab}\partial_a\partial_b~\delta^{(2)}(\xi-\zeta)=0.
\ee
The second functional derivatives of $\Gamma$ read then, in Fourier components (no summation on $i$) ,
\bea
\frac{\delta^2\Gamma}{\delta X^0(p)\delta X^0(q)}&=&
\frac{1}{4\pi}\left(2\kappa p^2+R^{(2)}\phi^{''}\right)\delta^2(p+q)\nn
\frac{\delta^2\Gamma}{\delta X^i(p)\delta X^i(q)}&=&
\frac{\tau_i ~p^2}{2\pi}\delta^2(p+q).
\eea

\noindent
The area of a sphere with curvature scalar $R^{(2)}$ is $8\pi/R^{(2)}$, so
with the constant configuration $X^\mu=x_\mu$ we have
\be\label{Gammaconfig}
\Gamma=2\phi_\lambda(x_0),
\ee
and the trace appearing in eq.(\ref{evolGappendix}) is
\bea\label{TrABA}
\frac{1}{4\pi}\mbox{Tr}\{\partial\partial (\delta^2\Gamma)^{-1}\}
&=&-\int\frac{d^2p}{(2\pi)^2}\left(\frac{p^2}{2\kappa p^2+R^{(2)}\phi^{''}}
+\frac{1}{2}\sum_{i=1}^{D-1}\frac{1}{\tau_i}\right)\frac{8\pi}{R^{(2)}}\nn
&=&-\frac{\Lambda^2}{R^{(2)}}\left(\frac{1}{\kappa}+\sum_{i=1}^{D-1}\frac{1}{\tau_i}\right)
+\frac{\phi^{''}}{2\kappa^2}\ln\left(1+\frac{2\Lambda^2\kappa}{R^{(2)}\phi^{''}}\right).
\eea
To compute this trace, we used the fact that
\be
\delta^2(p=0)=~\mbox{world-sheet area}~=\frac{8\pi}{R^{(2)}}.
\ee
The evolution equation for $\phi$
is finally obtained by putting together results
(\ref{evolGappendix}), (\ref{Gammaconfig}) and (\ref{TrABA}):
\be
\dot\phi=-\frac{\Lambda^2}{2R^{(2)}}\left(\frac{1}{\kappa}+\sum_{i=1}^{D-1}\frac{1}{\tau_i}\right)+
\frac{\phi^{''}}{4\kappa^2}\ln\left(1+\frac{2\Lambda^2\kappa}{R^{(2)}\phi^{''}}\right).
\ee
It is this equation we use in the current work in order to discuss non-trivial fixed point solutions ${\dot \phi} =0$ and their implications for String Cosmology.

\section*{Appendix B: Geometrical properties of the graviton
and dilaton backgrounds}

For the target-space metric $g_{\mu\nu}(X^0)=\mbox{diag}(\kappa(X^0),\tau_1(X^0), ...,\tau_{D-1}(X^0))$,
the non-vanishing components of the Christoffel symbol are (without summation over the space indice $i$)
\be
\Gamma^i_{~0i}=\frac{\tau_i^{'}}{2\tau_i},~~~~~~~
\Gamma^0_{~00}=\frac{\kappa^{'}}{2\kappa},~~~~~~~~
\Gamma^0_{~ii}=-\frac{\tau_i^{'}}{2\kappa},
\ee
so that the non vanishing covariant derivatives of the dilaton are
\bea
\nabla_0\nabla_0\phi&=&\phi^{''}-\frac{\kappa^{'}}{2\kappa}\phi^{'},\nn
\nabla_i\nabla_i\phi&=&\frac{\tau_i^{'}}{2\kappa}\phi^{'}.
\eea
The non-vanishing components of the Riemann and Ricci tensors are
\bea
R_{0i0}^{~~~~i}&=&-\left(\frac{\tau_i^{'}}{2\tau_i}\right)^{'}+\frac{\kappa^{'}\tau_i^{'}}{4\kappa\tau_i}
-\left(\frac{\tau_i^{'}}{2\tau_i}\right)^2  ,\nn
R_{iji}^{~~~~j}&=&-\frac{\tau_i^{'}\tau_j^{'}}{4\kappa\tau_j}\nn
R_{00}&=&\sum_{i=1}^{D-1}\left[-\left(\frac{\tau_i^{'}}{2\tau_i}\right)^{'}+\frac{\kappa^{'}\tau_i^{'}}{4\kappa\tau_i}
-\left(\frac{\tau_i^{'}}{2\tau_i}\right)^2\right]\nn
R_{ii}&=&-\frac{\tau_i^{'}}{4\kappa}\left( \sum_{j\ne i}\frac{\tau_j^{'}}{\tau_j}\right)
+\frac{\tau_i}{\kappa}\left[ -\left(\frac{\tau_i^{'}}{2\tau_i}\right)^{'}+\frac{\kappa^{'}\tau_i^{'}}{4\kappa\tau_i}
-\left(\frac{\tau_i^{'}}{2\tau_i}\right)^2\right]
\eea
Together with the condition $\kappa\propto\phi^{''}$, one can see that, with power-law dependence on $x^0$ of the metric components,
$R_{00}$ is homogeneous to $\nabla_0\nabla_0\phi$ only if
\bea
\phi(X^0)&=&\phi_0\ln(X^0)\nn
\kappa(X^0)&=&\frac{\kappa_0}{(X^0)^2},
\eea
as was the case in ref.~\cite{AEM1}, where the antisymmetric tensor field had not been taken into account.

\end{document}